\documentclass[11pt,twoside]{article}
\usepackage{asp2014}

\aspSuppressVolSlug
\resetcounters

\begin{document}

\title{Latent Space Explorer: Unsupervised Data Pattern Discovery on the Cloud}

\author{Thomas~Cecconello$^1$, Cristobal Bordiu$^2$, Filomena Bufano$^2$, Lucas Puerari$^1$, Simone Riggi$^2$, Eugenio Schisano$^3$, Eva Sciacca$^2$, Ylenia Maruccia$^3$, and Giuseppe Vizzari$^1$
\affil{$^1$University of Milano-Bicocca, Milano, IT; \\
$^2$Osservatorio Astronomico di Catania (OACT-INAF),
Catania, IT \\
$^3$Istituto di Astrofisica e Planetologia Spaziali (IAPS-INAF), Rome, IT\\
\email{thomas.cecconello@unimib.it}
}}

\paperauthor{Thomas~Cecconello}{thomas.cecconello@unimib.it}{0000-0001-6519-5011}{University of Milano-Bicocca}{Dipartimento di Informatica, Sistemistica e Comunicazione DISCo}{Milano}{IT}{20126}{Italy}
\paperauthor{Cristobal~Bordiu}{cristobal.bordiu@inaf.it}{0000-0002-7703-0692}{INAF}{Osservatorio Astrofisico di Catania}{Catania}{IT}{95123}{Italy}
\paperauthor{Filomena~Bufano}{filomena.bufano@inaf.it}{0000-0002-3429-2481}{INAF}{Osservatorio Astrofisico di Catania}{Catania}{IT}{95123}{Italy}
\paperauthor{Lucas~Puerari}{lucas.puerari@unimib.it}{}{University of Milano-Bicocca}{Dipartimento di Informatica, Sistemistica e Comunicazione DISCo}{Milano}{IT}{20126}{Italy}
\paperauthor{Simone~Riggi}{simone.riggi@inaf.it}{0000-0001-6368-8330}{INAF}{Osservatorio Astrofisico di Catania}{Catania}{IT}{95123}{Italy}
\paperauthor{Eva~Sciacca}{eva.sciacca@inaf.it}{0000-0002-5574-2787}{INAF}{Osservatorio Astrofisico di Catania}{Catania}{IT}{95123}{Italy}
\paperauthor{Eugenio~Schisano}{eugenio.schisano@inaf.it}{0000-0003-1560-3958}{INAF}{Istituto di Astrofisica e Planetologia Spaziali}{Roma}{IT}{00133}{Italy} 
\paperauthor{Giuseppe~Vizzari}{giuseppe.vizzari@unimib.it}{0000-0002-7916-6438}{University of Milano -- Bicocca}{Dipartimento di Informatica, Sistemistica e Comunicazione DISCo}{Milano}{IT}{20126}{Italy}



  
\begin{abstract}

The amount of data describing astronomical phenomena is growing at an overwhelming rate: tools supporting a computer assisted analysis of these data is of ever-growing importance. We propose an approach based on unsupervised machine learning as a central element of an overall workflow, involving domain experts and computer scientists; the workflow includes three phases: (i) achieving a compact representation of elements of the dataset by means of representation learning techniques, shifting the analysis from cumbersome representations to compact vectors in a latent space, (ii) visualizing results of this analysis in a 2D or 3D space (further reducing dimensionality of the space) and (iii) potentially clustering points associated to instances to suggest patterns to the domain experts that will evaluate their potential meaning within the domain. This work presents the overall approach within a cloud based setting, and results on two specific astronomical research topics, namely the study of galactic supernova remnants and star forming clumps.
  
\end{abstract}

\section{Introduction}
\label{sec:intro}
Extracting information from raw data is probably one of the central activities of experimental scientific enterprises. This work is about a pipeline in which a specific module is trained to provide a compact, essential representation of the training data, useful as a starting point for visualization and analyses aimed at detecting patterns, regularities among data. To enable researchers exploiting this approach, a cloud based system is being developed and tested in the NEANIAS project as one of the ML-tools of a thematic service to be offered to the EOSC. Here, we describe the architecture of the system and introduce two example use cases in the astronomical context.

\section{Proposed pipeline}
\label{sec:workflow}

\begin{figure}[btp]
    \centering
    \includegraphics[width=0.9\textwidth]{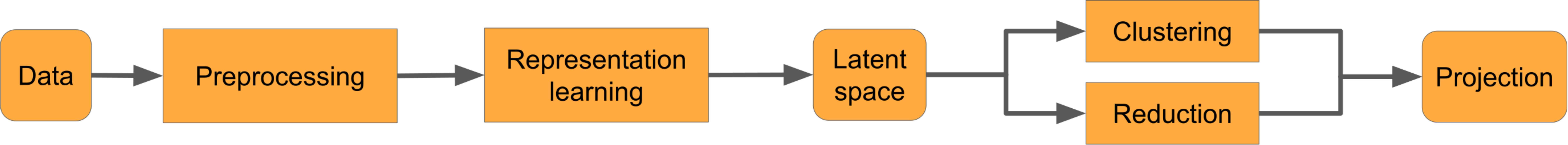}
    \caption{Proposed pipeline.}
    \label{fig:pipeline}
\end{figure}

Extracting information from dataset is a multistep process (Fig.\ref{fig:pipeline}) and starts with a \textbf{preprocessing} step: while this kind of activity is present even with everyday computer vision workflows, on astronomic imaging (especially if the sensor is not optical, such as a radio antenna) it needs to be tuned for each specific case. A major issue is to deal with NaN values, that are problematic for the representation learning step and thus must be replaced with meaningful values or ignored by the model.

The prepared data is then ingested by the \textbf{representation learning}  module \citep{jing2020self}. This task makes use of deep neural networks trained in a way that does not require manually annotated data (e.g. autoencoders, SimCLR); such networks essentially learn functions that map data into a features space that we'll call \textit{latent space}. Although more compact than the original one, this space is still multidimensional and so, to be visual analysed in 2D or 3D, \textbf{reduction} methods like PCA, T-SNE and UMAP need to be adopted. Part of the information about the placement in a multidimensional space is lost, so to add information about plausible patterns in the full dimensional space it is possible to execute \textbf{clustering} algorithms like k-means and DBSCAN. 

Results produced by the latest steps could be \textit{projected} in a 2D or 3D space. The \textit{latent space explorer} service let the user interact with that space and to download a list of clusters to be analysed in detail with proper tools.

\section{Cloud implementation}
\label{sec:ecosystem}
An implementation of this approach is going to be released as a cloud service within NEANIAS project and on the EOSC (European Open Science Cloud) portal. The pattern of a cloud application takes advantage of modularity, switching from monolithic pattern to microservices. This also simplifies exploiting core services developed in the NEANIAS overall ecosystem, like authentication, storage, logging and accounting. Furthermore, by isolating each microservice, it is possible to integrate the \textit{latent space explorer}\footnote{\url{https://lse.neanias.eu}} with multiple type of core services. The framework used to achieve the above described modularity is Kubernetes, relying on the isolation granted by docker containers. Kubernetes provides even the chance to scale the application based on some metrics like active users or actual computational load. In that way it is possible to serve an elastic amount of researchers, keeping high the quality of service.

\section{Use cases}
\label{sec:usecases}
\subsection{Star forming clumps}
\begin{figure}[tbp]
    \centering
    \includegraphics[width=0.9\textwidth,trim=0cm 6cm 0cm 1.3cm]{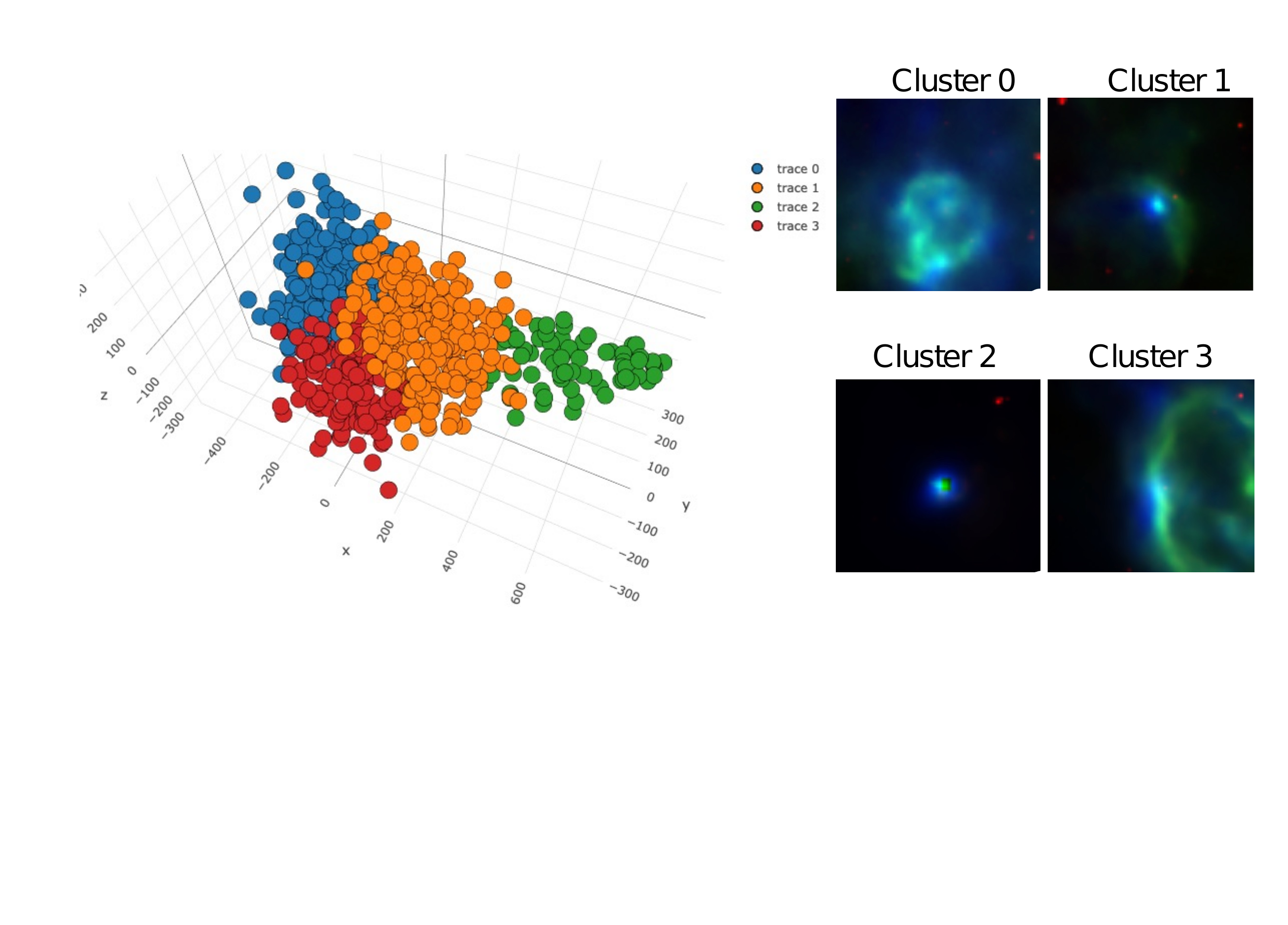}
    \caption{Pipeline output for the clumps use case (3D PCA, k-means, k=4).} 
    \label{fig:clumps}
\end{figure}

How stars form is one of the main questions in modern astrophysics. While the details of the formation process of low-mass stars are outlined, a lot of questions are still open on how  massive stars (M > 8 M$_\odot$) and star clusters form. Recent surveys like GLIMPSE, MIPSGAL, Hi-GAL, mapped large areas of our Galaxy in near, mid and far infrared, respectively, providing the data to understand how star formation takes place. The amount of data provided by these surveys is enormous: for example, more than 150000 objects, named clumps, were identified in Hi-GAL survey \citep{2021MNRAS.504.2742E}, while the number of objects identified by GLIMPSE and MIPSGAL  reaches several millions. A complete study on star formation requires the comparative analysis of all these data, both source catalogues and images from near-IR to radio frequencies, to identify clumps and hosted young protostars, to measure their properties from images and to classify and order them in terms of an evolutionary scheme, defining the star formation model. 
Currently, astrophysicists split the young objects into four classes, based on their appearance at different wavelengths, representing different stages of star formation. It is unclear if these classes are independent, if they are fully representative, or if there are missing classes in the description. Moreover, these classes do not include the  environmental effects that were recently found to set the initial condition for star formation, such as the filamentary shape of the surrounding cloud \citep{2020MNRAS.492.5420S}. 
Unsupervised learning techniques applied directly to the images have the potential to answer these and similar questions, and they can support scientists to handle the large datasets that are progressively becoming available. 
Here, we are applying for the first time these techniques to the ALMAGAL sample of more of 1000  dense clump identified by astrophysicists to be a representative of the entire early evolution path of protoclusters, from the prestellar to the ultracompact H\textsc{ii} region phase. Fig. \ref{fig:clumps} shows the results from the pipeline described in Section \ref{sec:workflow} to this sample, processing the images of ALMAGAL clumps covering a field of 3.5$\times$3.5 arcmin, to include regions of the surrounding cloud, in the wavelength range between 3 and 500 $\mu$m.




\subsection{SNRs}
Supernova Remnants (SNRs), the debris that remain after the final explosion of a star, are among the principal polluters in the Galaxy. Bounded by violent shockwaves, these expanding clouds of dust and gas slowly dissipate into the interstellar medium, releasing vast amounts of energy and pouring their surroundings with heavy elements. Consequently, SNRs profoundly impact the structure, dynamics, and chemical composition of the Milky Way, being essential for our understanding of galactic evolution. The galactic census of SNRs consists of roughly 300 objects detected so far \citep{2019JApA...40...36G}. Such a relatively scarce population constitutes a  rather heterogeneous sample, with objects that exhibit a wide variety of observational properties: SNRs morphologies include shells, bubbles, knots and filaments, often showing strong asymmetries; some of them are bright at infrared wavelengths owing to dust emission; others only exhibit thermal and non-thermal radio emission. These differences are likely determined by the nature of the progenitor star, existing inhomogeneities in the pre-SN circumstellar material, and explosion dynamics, all of them poorly constrained aspects in most cases. Therefore, an important question arises: \textit{are there any underlying patterns in the SNR population, allowing for linking their observational properties with the progenitor stars and their surroundings?} Unsupervised learning seems the right approach to tackle this question from a systematic perspective. 

\begin{figure}[tbp]
    \centering
    \includegraphics[width=0.9\textwidth]{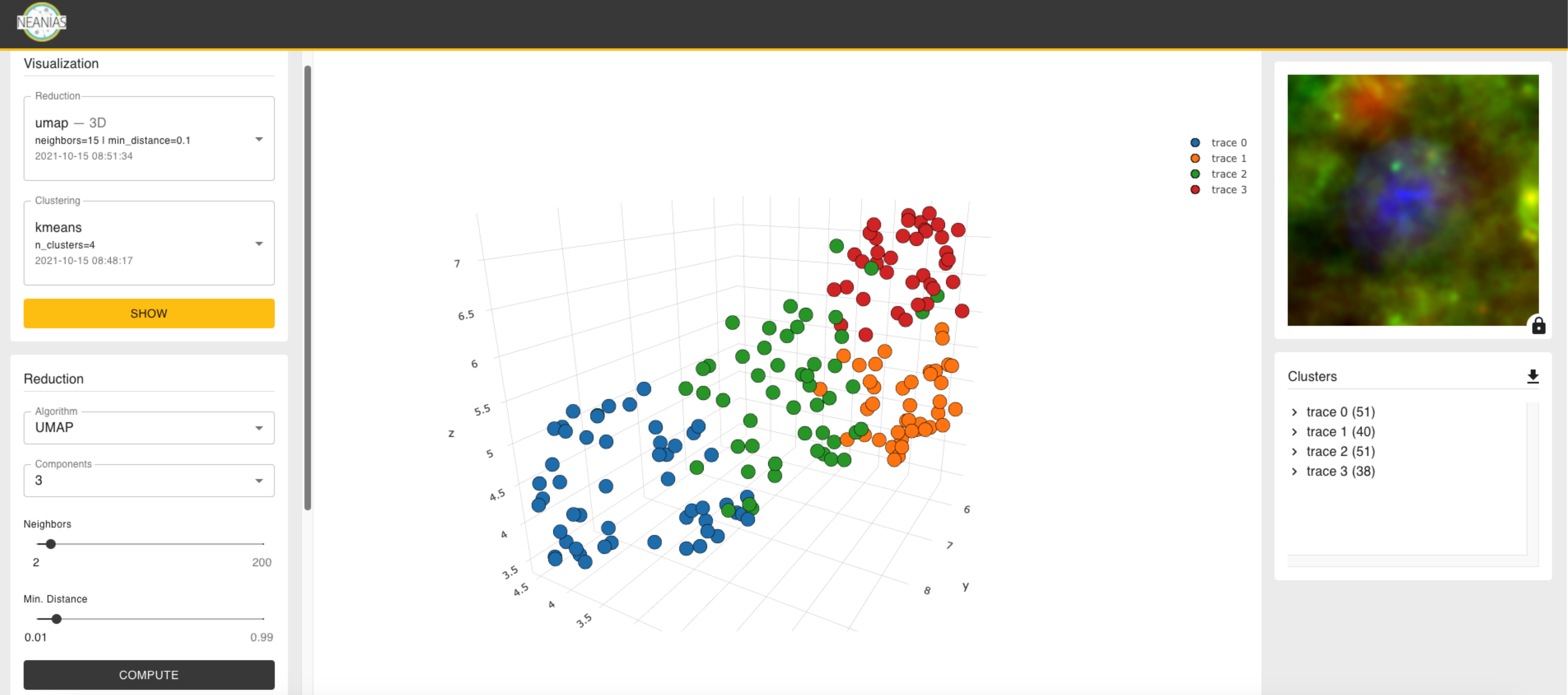}
    \caption{Pipeline output for the SNR use case (UMAP 3D, k-means, K=4).} 
    \label{fig:snr}
\end{figure}

We have collected multiwavelength imagery from different all-sky surveys, combining available mid-infrared (wavelength $\sim$22 $\mu$m), far-infrared (70 $\mu$m) and radio continuum (30 cm) data into a representative sample of 181 SNRs. Taking advantage of the pipeline described in Section \ref{sec:workflow}, we have explored different clustering strategies to unveil the underlying structure of the sample. Fig. \ref{fig:snr} shows an example output, with the results for a k-means clustering ($K=4$) represented on a 3D UMAP projection. While this is still a work in progress, and we are yet to provide a meaningful physical interpretation for the resulting clusters, the first results are encouraging.
\vspace{-0.5cm}

\section{Summary and future work}
In this work, we presented the workflow of a cloud based service employing machine learning techniques to assist astrocatalogue, discovering patterns in multiwavelength data. The service will soon be available in the NEANIAS catalogue allowing interested researchers to experiment it online.

\acknowledgements The research has received funding from the EC Horizon 2020 grant agreement No. 863448 (NEANIAS).

\bibliography{X0-019.bib}
\end{document}